\documentclass{mn2e}
\usepackage{graphicx}

\newcommand{\loc}{{\rm{}loc}}

\begin{document}
\title{Polarization signatures of strong gravity in AGN accretion discs}

\author[M.~Dov\v{c}iak, V.~Karas and G.~Matt]
{M.~Dov\v{c}iak,$^{\!1}$
 V.~Karas$^{1,2}$ and 
 G.~Matt$^{3}$ \\~\\
$^1$~Astronomical Institute, Academy of Sciences,
 Bo\v{c}n\'{\i}~II, CZ-140\,31~Prague, Czech Republic\\
$^2$~Faculty of Mathematics and Physics,~Charles University, 
 V~Hole\v{s}ovi\v{c}k\'ach~2, CZ-180\,00~Prague, Czech Republic\\
$^3$~Dipartimento di Fisica, Universit\`a degli Studi ``Roma Tre'', 
 Via della Vasca Navale 84, I-00146~Roma, Italy}

\date{Accepted .... Received ...}
\pagerange{\pageref{firstpage}--\pageref{lastpage}}
\pubyear{2004}
\maketitle
\label{firstpage}
\begin{abstract}
The effects of strong gravity on the polarization of the Compton
reflection from an X-ray illuminated accretion disc are studied.  The
gravitational field of a rotating black hole influences Stokes
parameters of the radiation along the propagation to a distant observer.
Assuming the lamp-post model, the degree and the angle of polarization
are examined as functions of the observer's inclination angle, of the
height of the primary source and of the inner radius of the disc
emitting region. It is shown that polarimetry can provide essential
information on the properties of black holes sources, and it is argued
that time variation of the polarization angle is a strong signature of
general relativity effects. The expected polarization degree and angle
should be detectable by new generation polarimeters, like the one
planned for the {\it{}Xeus} mission.

\end{abstract}


\section{Introduction}
Accretion discs in central regions of active galactic nuclei (AGN) are
subject to strong external illumination originating from some kind of
corona and giving rise to specific spectral features in the X-ray band,
namely a Compton Reflection component and fluorescent emission lines,
the most prominent of them being that of iron. It has been shown that
the shape of the intrinsic spectra must be further modified by the
strong gravitational field of the central mass, and so X-ray
spectroscopy could allow us to explore the innermost regions of
accretion flows near super-massive black holes (for recent reviews, see
Fabian et al.\ 2000; Reynolds \& Nowak 2003). Similar mechanisms operate
also in some Galactic black hole candidates.

However, a rather surprising result from recent XMM--{\it{}Newton}
observation is that relativistic iron lines are not as common as
previously believed (Bianchi et al.\ 2004 and references therein). This
does not necessarily mean that the iron line is not produced in the
innermost regions of accretion discs. The situation is likely more
complex than in simple, steady scenarios, and indeed some evidence for
the line emission arising from orbiting spots is present in the
time-resolved spectra of a few AGN (e.g., Dov\v{c}iak et al.\ 2004a, and
references therein). Even when clearly observed, relativistic lines
behave differently than expected. The best example is the puzzling lack
of correlation  between line and continuum emission in MCG--6-30-15
(Fabian et al.\ 2002), unexpected because the very broad line profile
clearly indicates that the line originates in the innermost regions of
the accretion disc, hence very close to the illuminating source. 
Miniutti et al.\ (2003, 2004) have proposed a solution to this problem
in terms of the source moving along the black hole rotation
axis or very close to it. 

In this paper we demonstrate that polarimetric studies could be very
useful to discriminate between different geometries and physical states
of accreting sources in strong gravity regime. The idea of using
polarimetry to gain additional information about compact objects is not
a new one. In this context it was proposed by Rees (1975) that polarized
X-rays are of high relevance. Pozdnyakov, Sobol \& Sunyaev (1979)
studied spectral profiles of iron X-ray lines that result from multiple
Compton scattering. {Laor et al.\ (1990) studied the polarization of a
thin accretion disc around a rotating black hole within the framework
of general relativity, using an improved radiative transfer computations.}
Later on, various influences affecting polarization
(due to magnetic fields, absorption as well as strong gravity) were
examined for black hole accretion discs (Agol \& Blaes 1996; Ogura, Ohuo \&
Kojima 2000). Temporal variations of polarization were also discussed,
in particular the case of orbiting spots near a black hole (Connors,
Piran \& Stark 1980; Bao, Wiita \& Hadrava 1996). Here we examine
consequences of a specific model of an illuminated accretion disc. 
With the promise of new polarimetric detectors (Costa et al.\ 2001), 
quantitative examination of specific models becomes timely. 

Since the reflecting medium has a disc-like geometry, a substantial
amount of linear polarization is expected in the resulting spectrum
because of Compton scattering. Polarization properties of the disc
emission are modified by the photon propagation in gravitational field,
providing additional information on its structure. In this paper we
calculate the observed polarization of the reflected radiation  assuming
the lamp-post model (Martocchia \& Matt
1996;  Petrucci \& Henri 1997; Ghisellini et al. 2004).  We assume a
rotating (Kerr) black hole as the only source of the  gravitational
field, having a common symmetry axis with an accretion disc. 

In the next section, by employing Monte-Carlo computations (Matt, Perola
\& Piro 1991; Matt 1993) we will find the intrinsic emissivity of an
illuminated disc. Then we will integrate contributions to the total
signal across the disc emitting region and, using a general relativistic
ray-tracing technique (Dov\v{c}iak et al.\ 2004b, 2004c), we derive the
Stokes parameters as measured by a distant observer. We will present the
polarization properties of scattered light as a function of model
parameters, namely, the height $z=h$ of the primary source on the
symmetry axis, the dimension-less angular momentum $a$ of the black
hole, and the viewing angle $\theta_{\rm{}o}$ of the observer. {The
conclusion will be that the X-ray polarimetry can be very useful to help
discriminating between models and determining their parameters.}

\section{Polarization of reflected light in the strong gravity regime}
\subsection{An irradiated disc near a black hole}
\label{lamp-post}
The local emissivity of radiation reflected at a given point of the disc
is assumed to be proportional to the incident illumination by the
primary power-law continuum. The primary source of the lamp-post model
is located at height $h$ above the black hole, and
so we first need to integrate light rays from the source down to the
disc. The disc is assumed to be stationary and we restrict ourselves to
the time-averaged analysis, assuming processes that
vary in a much slower pace than the light-crossing time at the
corresponding radius.

An incident photon strikes the disc at a certain point ${\cal P}(r,\varphi)$ 
in equatorial plane $\theta=\pi/2$ (i.e.\ $z=0$). Four-momentum of 
an incident photon is
\begin{eqnarray}
p_{\rm{}i}^t & = & 1+2r^{-1}+4\Delta^{-1}\, ,\\
p_{\rm{}i}^r & = & {\cal R}_{\sigma}\big[\left(r^2+a^2\right)^2
 -{\Delta}r^{-2}\left(a^2+q_{\rm{}p}^2\right)\big]^{1/2}\, ,\\
p_{\rm{}i}^\theta & = & q_{\rm{}p}r^{-2}\, ,\\
p_{\rm{}i}^\varphi & = & 2ar^{-1}\Delta^{-1}\, ,
\end{eqnarray}
where $q_{\rm{}p}^2\equiv\sin^2{\!\theta_{\rm{}p}}
(h^2+a^2)^2\Delta_{\rm{}p}^{-1}-a^2$ is Carter's constant of motion,
$\Delta(r){\equiv}r^2-2r+a^2$, $\Delta_{\rm{}p}{\equiv}\Delta(h)$, and
${\cal R}_{\sigma}$ is the sign of $p_{\rm{}i}^r$. Usual notation is
employed for Kerr metric functions in Boyer-Lindquist coordinates 
(e.g.\ Kato, Fukue \& Mineshige 1998).  All quantities
are made dimensionless by scaling with gravitational radius $GM/c^2$.
The disc is defined by $r_{\rm{}in}{\leq}r{\leq}r_{\rm{}out}$.  
We also denoted $\theta_{\rm{}p}$ to be the local angle
of emission under which photon emerges in the source rest frame 
($\theta_{\rm{}p}=0$ corresponds to a photon heading downwards to the
disc, while $\theta_{\rm{}p}=180^\circ$ is for upward direction). Affine
parametrization has been employed in such a way that  conserved energy
of an incident photon is  $-p_{{\rm{}i}\,t}=-p_{{\rm{}p}\,t}=1$ and its
conserved angular momentum  vanishes, i.e.\ $l_{\rm{}p}=0$. The assumed
range of key parameters is $0{\leq}a{\leq}1$,
$0{\leq}\theta_{\rm{}o}{\leq}90^{\circ}$ (in this paper we show results
for a rapidly rotating black hole, $a\rightarrow1$, and we use $5^\circ$
resolution in $\theta_{\rm{}o}$). 

The following considerations are necessary in order to derive the observed
spectrum and polarization. First, the gravitational and Doppler-induced 
shift of energy of the primary photons impinging on the disc is
\begin{equation}
\label{gfac_lamp}
g_{\rm{}p}= \frac{E_{\rm{}i}}{E_{\rm{}p}}=
\frac{p_{{\rm{}i}\,\mu} U^\mu}{p_{{\rm{}p}\,\alpha} U_{\rm{}p}^\alpha}=
-\frac{p_{{\rm{}i}\,\mu} U^\mu}{U_{\rm{}p}^t}\,,
\end{equation}
where $E_{\rm{}p}$ and $E_{\rm{}i}$ denote the photon energy	
at the point of emission from the primary source and at the 
point of incidence on the disc, respectively. $U^\mu$ is Keplerian
four-velocity  of the disc medium and $U_{\rm{}p}^\alpha$ is
four-velocity of the primary  source. The only non-zero component of the
latter quantity is  $U_{\rm{}p}^t=\sqrt{(h^2+a^2)\Delta_{\rm{}p}^{-1}}$.

Cosine of local incidence angle is
\begin{equation}
\label{cosine_inc}
\mu_{\rm{}i}=|\cos{\delta_{\rm{}i}}\,|=
\displaystyle\frac{{p_{{\rm{}i}\,\alpha}\,n^{\alpha}}}
{{p_{{\rm{}i}\,\mu}\,U^{\mu}}}\, ,
\end{equation}
where $n^\alpha=-e_{(\theta)}^\alpha=(0,0,-r^{-1},0)$ denotes normal 
direction to the disc in the co-moving frame of the disc medium. For 
explicit definition of the local tetrad $e_{(a)}^\alpha$ and further
details, see Dov\v{c}iak et al.\ (2004b).

The reflection component has been computed 
by a Monte-Carlo code (Matt 1993). The number of reflected photons is
proportional to the incident flux $N_{\rm{}i}^{S}(E_{\rm{}p})$
arriving from the primary source,
\begin{equation}
\label{illumination}
N_{\rm{}i}^{S}(E_{\rm{}p})=N_{\rm{}p}^{\Omega}(E_{\rm{}p})\frac{{\rm{}d}
\Omega_{\rm{}p}}{{\rm{}d}S_{\loc}}\,,
\end{equation}
where $N_{\rm{}p}^{\Omega}(E_{\rm{}p})=N_{0
{\rm{}p}}\,E_{\rm{}p}^{-\Gamma}$  represents an isotropic and steady
power-law primary emission that is emitted into solid angle
${\rm{}d}\Omega_{\rm{}p}$ and eventually illuminates the local
area element ${\rm{}d}S_{\loc}$ on the disc.  The ratio
${\rm{}d}\Omega_{\rm{}p}/{\rm{}d}S_{\loc}$ is
\begin{equation}
\frac{{\rm{}d}\Omega_{\rm{}p}}{{\rm{}d}S_{\loc}}=\frac{{\rm{}d}\Omega_{\rm{}p}}
{{\rm{}d}S} \frac{{\rm{}d}S}{{\rm{}d}S_\loc}=\frac{\sin{\theta_{\rm{}p}{\rm{}d}
\theta_{\rm{}p}}}{{\rm{}d}r}\frac{{\rm{}d}S}
{{\rm{}d}S_\loc}\, ,
\end{equation}
where ${\rm{}d}S=r^{-2}(p_{\rm{}i}^\theta)^{-1}\,{\rm{}d}S_\perp$ is the
element of coordinate area corresponding to proper area
${\rm{}d}S_\perp$ perpendicular to the incident ray. The proper
area projected onto the disc is
\begin{equation}
\nonumber
{\rm{}d}S_{\loc} =
\frac{U_{\rm{}p}^t}{rp_{\rm{}i}^\theta}\,g_{\rm{}p}\,
{\rm{}d}S_\perp\, .
\label{dSl}
\end{equation}
Only direct photon rays are considered in the present computations,
indirect image photons being known to produce merely a marginal
correction to the total signal under usual circumstances, {i.e., a
moderate inclination angle of the source. This is because luminosity of
the successive images of the same source decreases exponentially with
the order of the corresponding image (e.g., Luminet 1979).  The
contribution of indirect photons may be important only in case of a
highly inclined (edge-on) disc system,  in which case caustics play a
role (Bao et al.\ 1994; Rauch \& Blandford 1994).}  It follows from
equations (\ref{illumination})--(\ref{dSl}) that the  incident spectrum
on the disc conforms to a power-law profile in energy  with the same
photon index $\Gamma$ as the original primary  emission, i.e.\
$N_{\rm{}i}^{S}(E)=N_{0 {\rm{}i}}\,E^{-\Gamma}$ with the normalization
factor
\begin{equation}
N_{0 {\rm{}i}}=N_{0 {\rm{}p}}\,g_{\rm{}p}^{\Gamma-1}
\left(1-\frac{2h}{h^2+a^2}\right)^{1/2}\,\frac{\sin{\theta_{\rm{}p}}\,
{\rm{}d}\theta_{\rm{}p}}{r\,{\rm{}d}r}\, .
\end{equation}

\begin{figure*}
\includegraphics*[width=0.49\textwidth]{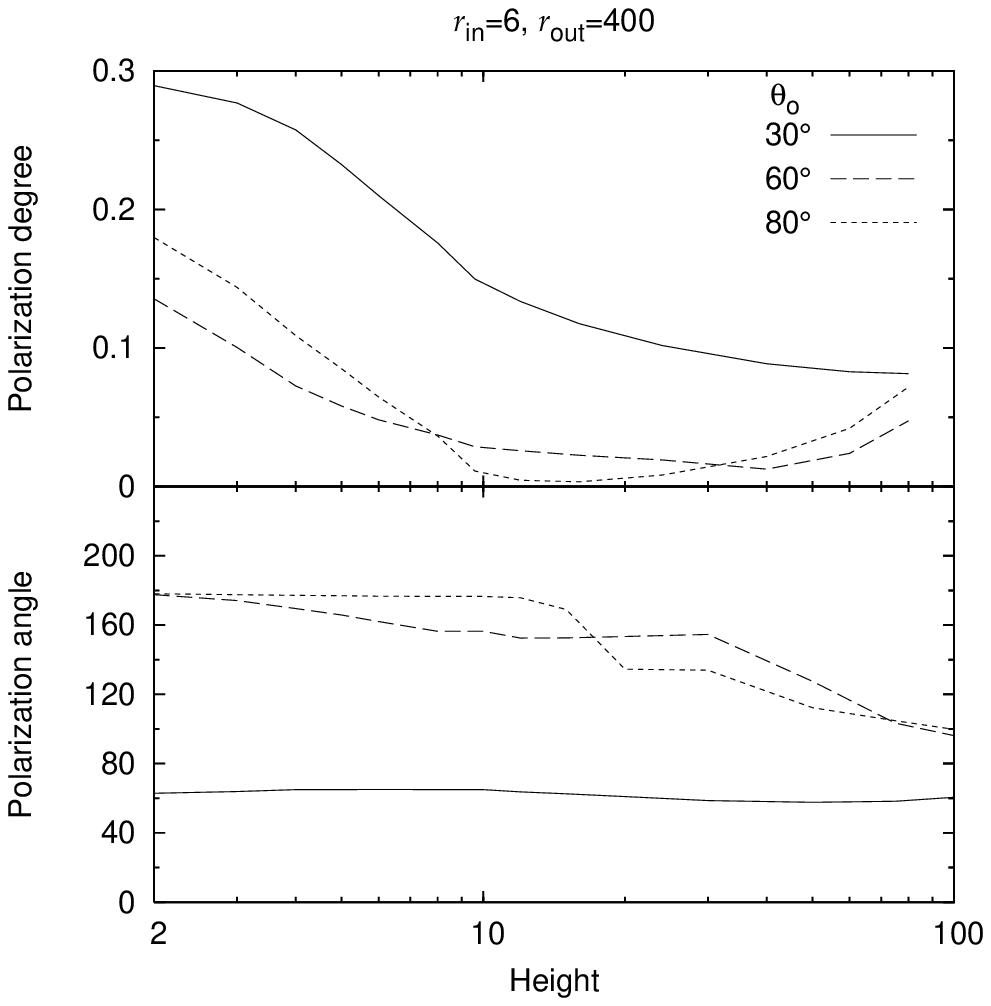}
\hfill
\includegraphics*[width=0.49\textwidth]{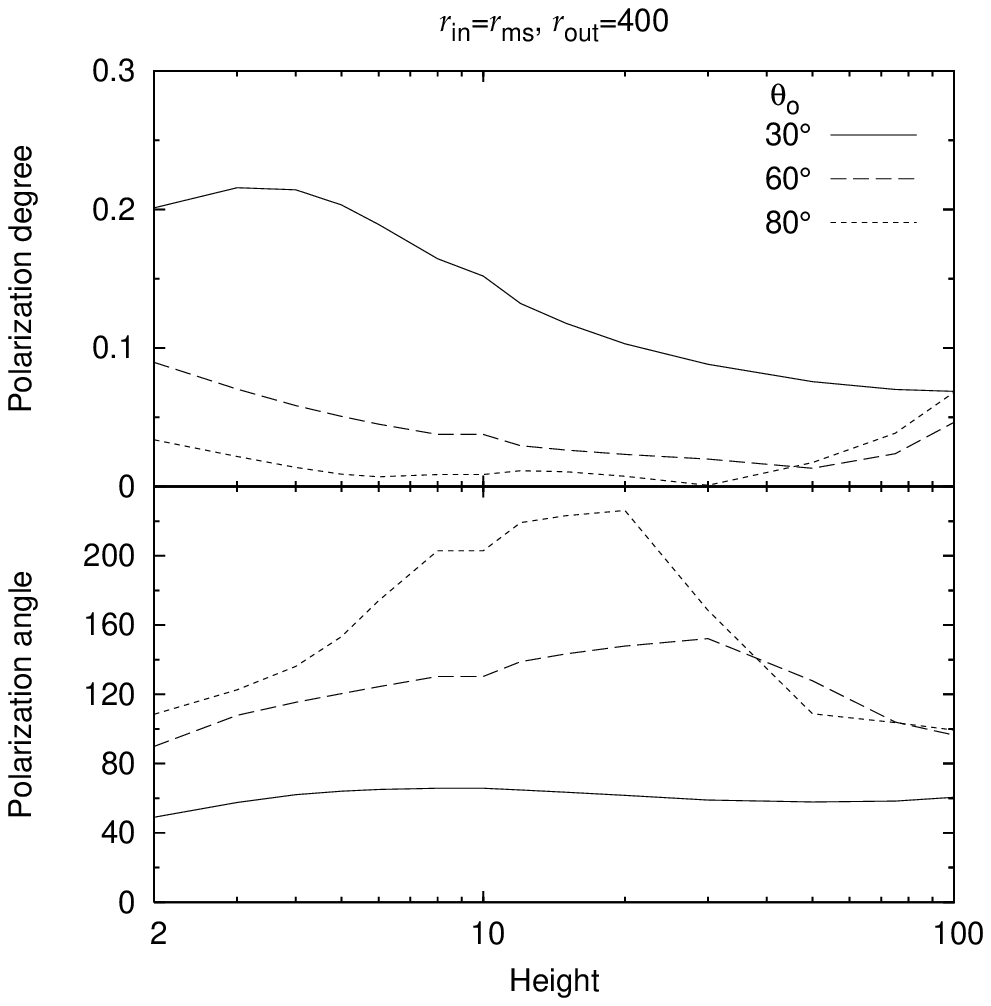}
\caption{Polarization degree and angle due to reflected radiation integrated 
over the whole surface of the disc and propagated to the point of
observation. The dependence on height $h$ is plotted for three values of
observer's view angle $\theta_{\rm{}o}$ (edge-on view of the disc
corresponds to $\theta_{\rm{}o}=90^\circ$). Left panel: $r_{\rm{}in}=6$;
right panel: $r_{\rm{}in}=1.20$. In both panels the energy range was
assumed $9$--$12$~keV, photon index of the  incident radiation
$\Gamma=2$, angular momentum $a=0.999$.}
\label{poldeg}
\end{figure*}

\begin{figure*}
\includegraphics*[width=0.48\textwidth]{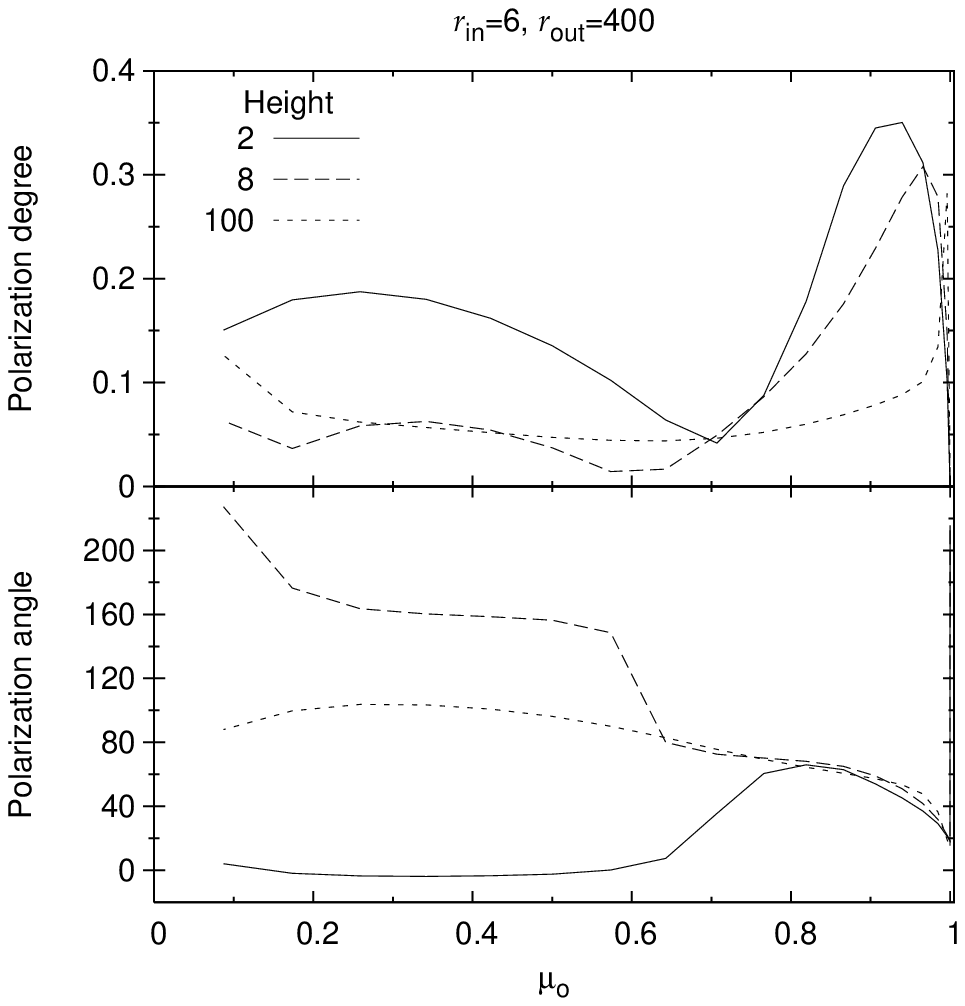}
\hfill
\includegraphics*[width=0.48\textwidth]{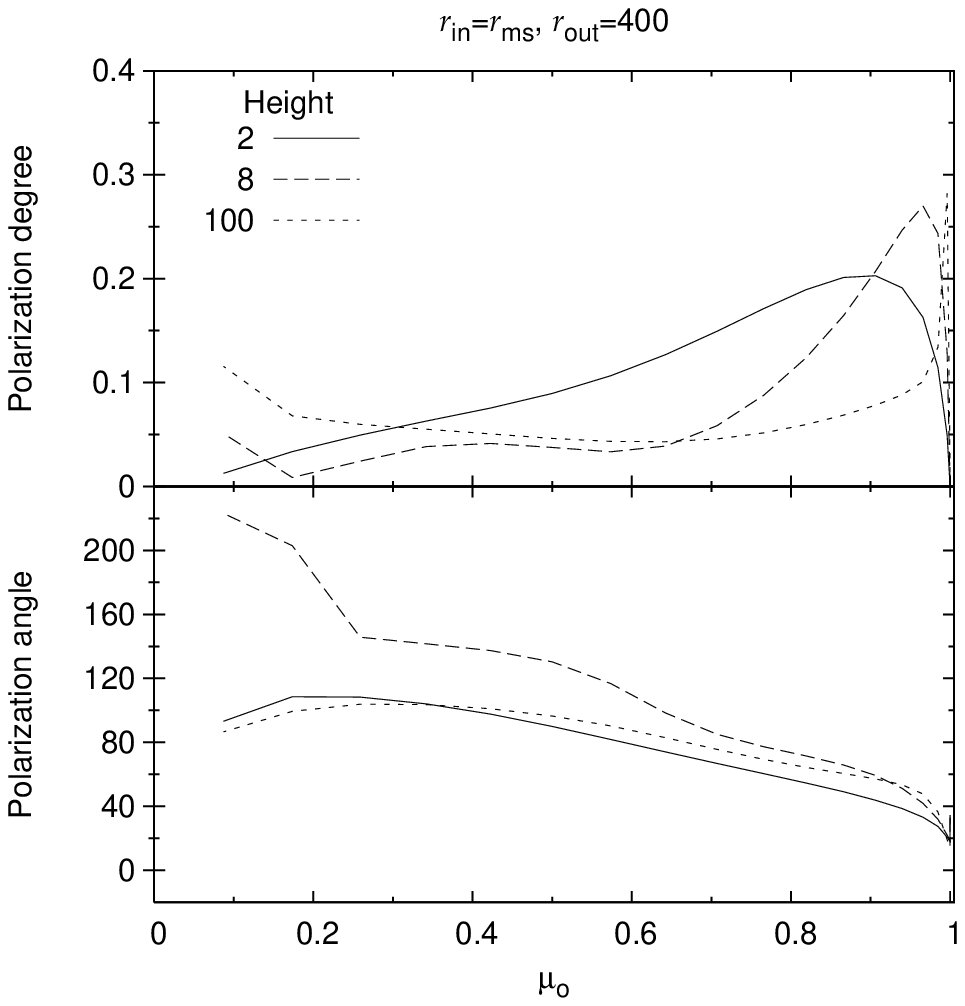}
\caption{Polarization degree and angle as functions of 
$\mu_{\rm{}o}$ (cosine of observer inclination) for the same model
in previous figure.}
\label{polangle1}
\end{figure*}

\subsection{Gravity induced changes of polarization}
The assumption of a geometrically thin accretion disc near a rotating
black hole defines geometry of the system and metric of the space-time. 
The intrinsic polarization of emerging light can be
then computed locally, assuming a plane-parallel scattering layer. This
problem was studied extensively in various approximations 
(Chandrasekhar 1960; Sunyaev \& Titarchuk 1985). Here we adopt the
approach of Matt (1993) and we refer the reader to recent papers
(Dov\v{c}iak et al.\ 2004b,\,c) for more details on the treatment of
polarized light in strong gravitational fields.

Four Stokes parameters, $I_{\nu}$, $Q_{\nu}$, $U_{\nu}$ and $V_{\nu}$, 
entirely describe polarization properties of the scattered light. 
Hereafter we will distinguish the quantities that are determined
locally at the point of emission on the disc surface (denoted by index
`loc') and those relevant to a distant observer (index `o').
We introduce specific Stokes parameters,
\begin{equation}
i_\nu\equiv\frac{I_{\nu}}{E}\,,\quad q_\nu\equiv\frac{Q_{\nu}}{E}\,,\quad
u_\nu\equiv\frac{U_{\nu}}{E}\,,\quad v_\nu\equiv\frac{V_{\nu}}{E}\,,
\end{equation}
and then specific Stokes parameters per energy bin, i.e.\ $\Delta
i_{\rm{}o}$, $\Delta q_{\rm{}o}$, $\Delta u_{\rm{}o}$ and $\Delta
v_{\rm{}o}$. The latter quantities are directly measurable, 
specifying the fluxes of photons with a given polarization. 
We can write
\begin{eqnarray}
\label{S1}
{\Delta}i_{\rm{}o}(E,\Delta E) & = & N_0\int{\rm{}d}S\,\int{\rm{}d}E_{\loc}\,
i_{\loc}(E_{\loc})\,F\, ,\\
\label{S2}
{\Delta}q_{\rm{}o}(E,\Delta E) & = & N_0\int{\rm{}d}S\,\int{\rm{}d}E_{\loc}\,
[q_{\loc}(E_{\loc})\cos{2\Psi} \nonumber \\
&& -u_{\loc}(E_{\loc})\sin{2\Psi}]\,F\, ,\\
\label{S3}
{\Delta}u_{\rm{}o}(E,\Delta E) & = & N_0\int{\rm{}d}S\,\int{\rm{}d}E_{\loc}\,
[q_{\loc}(E_{\loc})\sin{2\Psi} \nonumber \\
&& +u_{\loc}(E_{\loc})\cos{2\Psi}]\,F\, ,\\
\label{S4}
{\Delta}v_{\rm{}o}(E,\Delta E) & = & N_0\int{\rm{}d}S\,\int{\rm{}d}E_{\loc}\,
v_{\loc}(E_{\loc})\,F\, .
\end{eqnarray}
Here, $F\equiv F(r,\varphi)=g^2\,l\,\mu_{\rm{}e}\,r$ is the transfer
function, $g$ being the total energy shift between observed and emitted
photons, $l$ the lensing effect, $\mu_{\rm{}e}$ the cosine of the
emission angle, and $\Psi$ the angle by which a vector rotates while it
is parallelly transported along the light geodesic. One can refer to
$\Psi$ as the gravity-induced change of polarization angle, because
polarization vector is parallelly transported along the light ray. The
integration domain covers the X-rays emitting surface of the disc,
$r_{\rm{}in}{\leq}r{\leq}r_{\rm{}out}$. Notice that the local specific
Stokes parameters may depend on $r$, $\varphi$ and $\mu_{\rm{}e}$, which
we do not state explicitly in eqs.\ (\ref{S1})--(\ref{S4}) for
simplicity but we take this dependency into account when integrating
contributions to the observed values.

\begin{figure*}
\includegraphics*[width=0.48\textwidth]{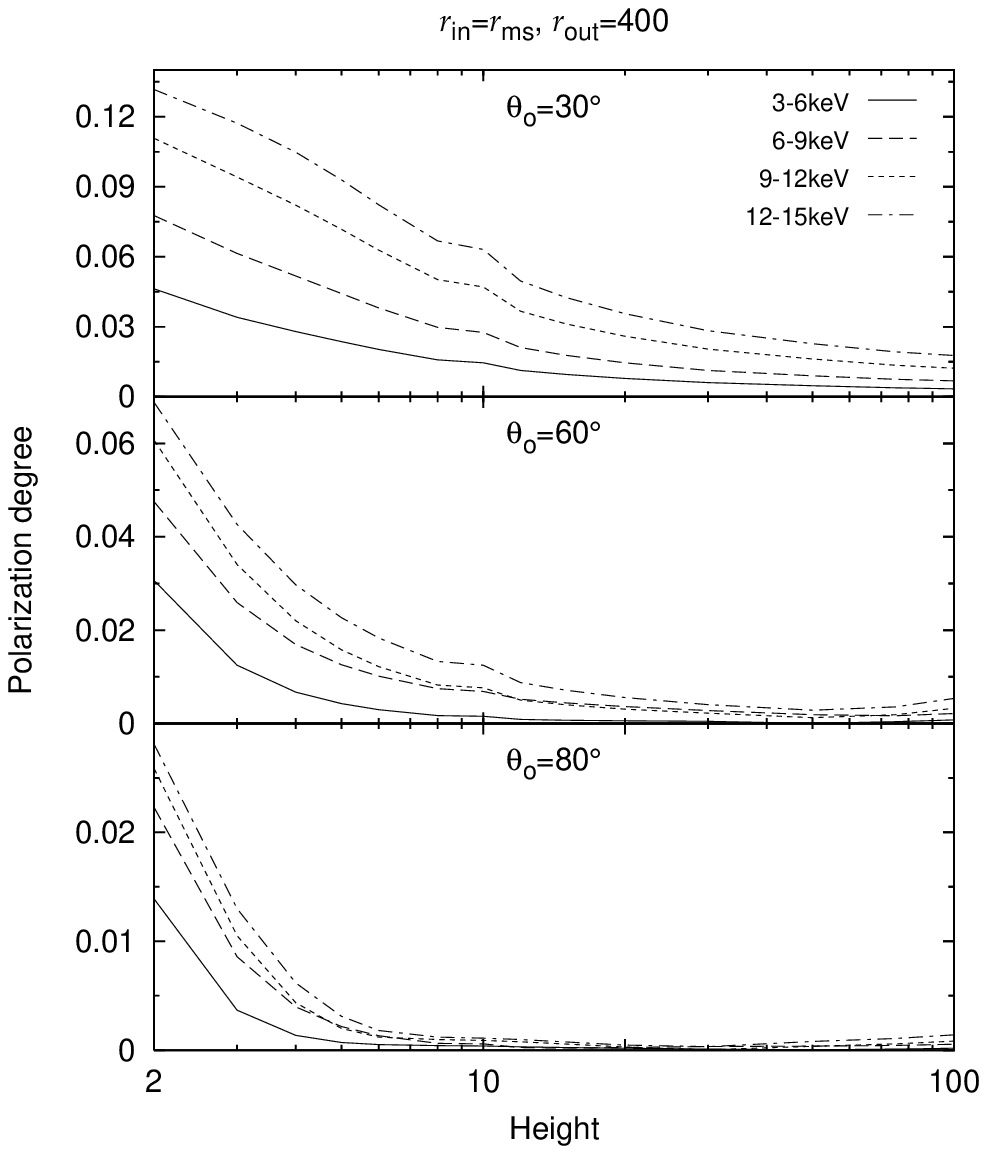}
\hfill
\includegraphics*[width=0.48\textwidth]{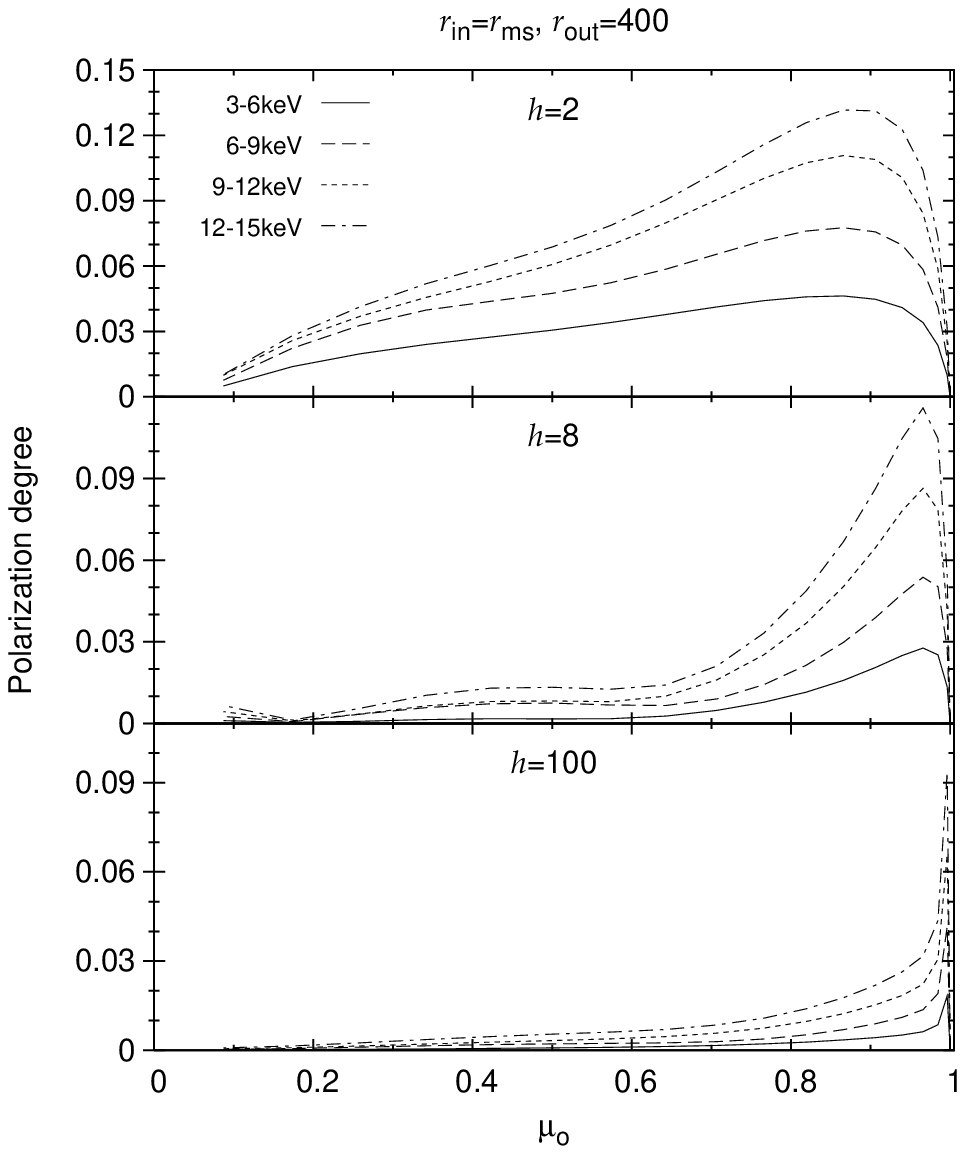}
\caption{Net polarization degree of the total (primary
plus reflected) signal as a function of height $h$ (left panel) and a
function of $\mu_{\rm{}o}$ (right panel). In each panel, curves are
parametrized by the corresponding energy range.}
\label{poldeg1}
\end{figure*}

A complementary way of characterizing polarization is in terms of
degree of polarization $P_{\rm{}o}$ and by two polarization angles
$\chi_{\rm{}o}$ and $\xi_{\rm{}o}$:
\begin{eqnarray}
P_{\rm{}o} & = & \sqrt{q_{\rm{}o}^2+u_{\rm{}o}^2+v_{\rm{}o}^2}/i_{\rm{}o}\,,\\
\tan{2\chi_{\rm{}o}} & = & u_{\rm{}o}/q_{\rm{}o}\, ,\\
\sin{2\xi_{\rm{}o}} & = & v_{\rm{}o}/\sqrt{q_{\rm{}o}^2+u_{\rm{}o}^2
+v_{\rm{}o}^2}\,.
\end{eqnarray} 
Unpolarized primary radiation was assumed and the approximation of
single Rayleigh scattering was adopted in our examples hereby. The
angular dependence of the local Stokes parameters is therefore given by
eqs.\ (I.147)  and (X.172) in Chandrasekhar (1960). See Matt (1993) and
Dov\v{c}iak et al.\ (2004b) for discussion of the  adopted approximation
in the present context. The angle $\xi_{\rm{}o}$  characterizes the
circular polarization and it vanishes in our case.

An explicit formula can be given for the angle $\Phi_{\rm{}i}$ between
the projection of three-momentum of an incident photon (in the local
rest-frame co-moving with the disc) and the radial vector,
\begin{equation}
\label{azim_angle_inc}
\Phi_{\rm{}i}=-{\cal R}_{\sigma}^{\rm{}i}\arccos\left(
\frac{-1}{\sqrt{1-\mu_{\rm{}i}^2}}\frac{\,{p_{{\rm{}i}\,\alpha}}
\,e_{(\varphi)}^{\alpha}}
{p_{{\rm{}i}\,\mu}U^\mu}\right)+\frac{\pi}{2}\, ,
\label{Phi}
\end{equation}
where ${\cal R}_{\sigma}^{\rm{}i}$ is positive for incident photons
travelling outwards ($p_{\rm{}i}^{(r)}>0$) and it is negative in case 
of inward direction ($p_{\rm{}i}^{(r)}<0$). The $\varphi$-component of
the local tetrad is
$e_{(\varphi)\,\alpha}=\sqrt{\Delta}(-U^{\varphi},0,0,U^t)$.
Equation~(\ref{Phi}) appears in evaluation of local Stokes parameters
(Chandrasekhar 1960).

Polarization of scattered light is shown in Figure~\ref{poldeg} where we
plot the polarization degree and the change of polarization angle as
functions of $h$ (we used a $14$ points grid with $h\leq100$). 
Notice that in the Newtonian case only polarization
angles of $0^{\circ}$ or $90^{\circ}$ would be expected for symmetry
reasons. The two panels in the figure correspond to different locations
of the inner disc edge: $r_{\rm{}in}=6$ and  $r_{\rm{}in}=1.20$,
respectively. The curves are strongly sensitive to $r_{\rm{}in}$ and
$h$, while the dependence on $r_{\rm{}out}$ is weak for a large disc
(here $r_{\rm{}out}=400$). Sensitivity to $r_{\rm{}in}$ is particularly
appealing if one remembers practical difficulties in estimating 
$r_{\rm{}in}$ by fitting spectra. The effect is clearly visible even 
for $h\sim20$. Graphs corresponding to $r_{\rm{}in}=6$ and $a=0.999$,
resemble, in essence quite closely, the non-rotating case ($a=0$)
because dragging effects are most prominent near horizon.

Figure~\ref{polangle1} shows the polarization degree and angle as
functions of the observer's inclination. Again, by comparing the two
cases of different $r_{\rm{}in}$ one can appreciate how sensitive
polarization is to details of the flow near the inner disc boundary.
Furthermore, it is worth noticing that the polarization degree has a
local maximum for moderate inclination angles ($\mu_{\rm{}o}\sim0.9$).
This peak occurs mainly due to interplay between the intrinsic
polarization on the disc and special-relativistic aberration of the
outgoing photons, while general relativistic effects are shaping details
of the profile.

The above-described polarization features are representative of the
scattering mechanism and the gravitational field structure acting on
reflected photons.  However, in order to compute directly observable
characteristics one has to combine  the primary continuum with the
reflected component. Polarization degree of the resulting signal depends
on mutual proportion of the two components and also on the energy range
of observation. In Figure~\ref{poldeg1}, we assumed that the irradiating
source emits isotropically and its light is affected only by
gravitational redshift and lensing, according to the source location at
$z=h$ on axis. This results in dilution of primary light by factor 
${\sim}g_{\rm{}h}^2(h,\theta_{\rm{}o})\,l_{\rm{}h}(h,\theta_{\rm{}o})$,
where $g_{\rm{}h}=\sqrt{1-2h/(a^2+h^2)}$ is the redshift of primary 
photons reaching directly the observer. The whole term is a combination
of special-relativistic transformation and general-relativity light
bending (factor $l_{\rm{}h}$ is order of unity and can be safely ignored
here; cf.\ Dov\v{c}iak et al.\ 2004b). Anisotropy of primary radiation
may further attenuate or amplify the polarization degree of the final
signal, while the polarization angle is rather independent of this
influence as long as primary light is itself unpolarized.

\section{Conclusions}  
We examined the polarimetric properties of X-ray illuminated accretion
discs in the lamp-post model. We found that observed values of Stokes
parameters are expected to be rather sensitive of the model parameters.
The adopted approach provides additional information with respect to
traditional spectroscopy, and so it has a great potential for
discriminating between different models. It offers an improved way of
measuring rotation of the black hole, because the radiation properties
of the inner disc region most likely reflect the value of the black hole
angular momentum. {Our calculations are complementary to those of
Laor et al.\ (1990), where the general relativity induced change of
polarization angle and degree were computed for the disc thermal
emission.}

While our calculations have been performed assuming a stationary
situation, in reality it is likely that the height of the illuminating
source changes with time, and indeed such variations have been invoked
by Miniutti et al.\ (2003) to explain the primary and reflected
variability patterns of MCG--6-30-15. Complete time-resolved analysis
(including all consequences of the light travel time in curved
spacetime) is beyond the scope of this paper and we defer it to future
work, assuming here that the primary source varies on a time-scale
longer than the light-crossing time in the system. This is a
well-substantiated assumption also from a practical point of view, since
feasible techniques will anyway require sufficient integration time
(i.e.\ of the order of several ksec). Here, suffice it to note that a
variation of $h$ implies a variation of the observed polarization angle
of the reflected radiation. {Although the change of polarization
angle and the degree of polarization are sensitive to general relativity
effects, to certain degree they depend also on other assumptions about
the source geometry and local physics of the source. However, it should
be obvious that time-resolved polarimetry can set constraints on the
models that are substantially more stringent than what can be achieved
by pure spectroscopy.}

New generation photoelectric polarimeters (Costa et al.\ 2001) in the
focal plane of large area optics (such as those foreseen for
{\it{}Xeus}) can probe polarization degree of the order of one percent 
in bright AGN, making polarimetry, along with timing and spectroscopy, a
tool for exploring the properties of the accretion flows in the vicinity
of black holes.

\section*{Acknowledgments}
VK and MD gratefully acknowledge support from Czech Science Foundation
grants 202/02/0735 and 205/03/0902, and from Charles University in
Prague ({\sc gauk} 299/2004). GM acknowledges financial support from
Agenzia Spaziale Italiana (ASI) and Ministero dell'Istruzione,
dell'Universit\`a e della Ricerca (MIUR), under grant {\sc
cofin--03--02--23}. 

{}
\label{lastpage}
\end{document}